\documentstyle[12pt]{article}
\documentstyle{fleqn}
\begin{document}
\begin{flushright}
hep-th/9403160 \\April 1994
\end{flushright}
\vspace{0.5in}
\begin{center}\Large{\bf Addendum to \lq\lq Classical and Quantum
Evolutions of the de Sitter and the anti-de Sitter Universes
in 2+1 dimensions" }(revised version)\\
\vspace{1cm}\renewcommand{\thefootnote}{\fnsymbol{footnote}}
\normalsize\ Kiyoshi Ezawa\footnote[1]{e-mail address:
ezawa@oskth.kek.jp }        \setcounter{footnote}{0}
\vspace{0.5in}

        Department of Physics \\
        Osaka University, Toyonaka, Osaka 560, Japan\\
\vspace{0.1in}
\end{center}
\vspace{0.8in}
\begin{abstract}
\baselineskip 20pt

The previous discussion \cite{ezawa} on reducing the phase space of
the first order Einstein gravity in 2+1 dimensions is reconsidered.
We construct a \lq\lq correct" physical phase
space in the case of positive cosmological constant,
taking into account the geometrical feature of SO(3,1) connections.
A parametrization which unifies the
two sectors of the physical phase space is also given.

\end{abstract}
\newpage

\baselineskip 20pt

In the previous paper \cite{ezawa} we have seen that, in 2+1 gravity
on ${\bf R}\times T^{2}$, the two sectors of the phase space of
Witten's Chern-Simons formulation (CSG) are related to the spaces of
solutions of the equations of motion in the ADM formalism when the
cosmological constant $\Lambda$ is positive.
We have used, however, a few manipulations which are mathematically
incorrect. Using the universal covering $\widetilde{\rm SO}(3,1)$ is
one of such manipulations. Since SL(2,{\bf C}) is homeomorphic to
${\bf R}^{3}\times S^{3}$ which is simply connected
\footnote{Roughly speaking, $S^{3}$ and ${\bf R}^{3}$ parametrize
respectively spatial rotations and boosts.}, it is
the universal covering of ${\rm SO}(3,1)_{0}$ which is the identity
component of SO(3,1). The use of $\widetilde{\rm SO}(3,1)$ therefore
does not allow us to distinguish the $4\pi$ differences in the
variables $u$ and $v$ which parametrize the standard sector
${\cal M}_{S}$. Nevertheless, the result obtained in \S\S 3.2 of
ref.\cite{ezawa} is physically adequate because a point on
${\cal M}_{S}$ specifies a unique spacetime constructed via the ADM.

The main purpose of this report is to give somewhat mathematically
improved construction of the physical phase space of CSG.

Let us recall the general case. It is known that the reduced phase
space $\check{\cal M}$ of the SO(3,1) Chern-Simons gauge theory
on ${\bf R}\times\Sigma$ equals the
moduli space of flat $SO(3,1)$ connections modulo gauge
transformations \cite{witte}.
This $\check{\cal M}$ is expected to be parametrized by
SL(2,{\bf C}) holonomy maps and therefore expected to be identified
with the moduli space of holonomy maps:
\begin{equation}
\check{\cal M}=Hom(\pi_{1}(\Sigma),{\rm SL(2,{\bf C})})/
\sim, \label{eq:CSGPS}
\end{equation}
where $\sim$ denotes the equivalence under the SL(2,{\bf C})
conjugations.

As an illustration let us consider the special case where the
spatial hypersurface has the topology of a torus $T^{2}$.
Since $\pi_{1}(T^{2})
\cong{\bf Z}\oplus{\bf Z}$, with two commutative generators $\alpha$
and $\beta$, holonomy maps are generated by two commuting elements
$S[\alpha]$ and $S[\beta]$ of SL(2,{\bf C}). Taking a proper
conjugation we have the following sectors of $\check{\cal M}$:
The \lq\lq standard sector" $\check{\cal M}_{S}$
\begin{equation}\left.  \begin{array}{l}
S[\alpha]=\exp(\frac{\sigma_{1}}{2i}(u+i\alpha)) \\
S[\beta]=\exp(\frac{\sigma_{1}}{2i}(v+i\beta))
\end{array} \right\}\quad with \quad u,v\in[-2\pi,2\pi),
\quad\alpha,\beta\in{\bf R};  \label{eq:HOSS}
\end{equation}
and the \lq\lq flat sector" $\check{\cal M}_{F}^{n_{1}n_{2}}$
\footnote{In ref.\cite{ezawa} this sector has been called the \lq\lq
null sector" ${\cal M}_{N}$. We no longer use ${\cal M}_{N}$ in
order to avoid confusing this sector with the null sector
${\cal M}_{n}$ in the $\Lambda=0$ case \cite{louko}.}
\begin{equation}\left.  \begin{array}{l}
S[\alpha]=(-)^{n_{1}}\exp(\frac{\eta}{2i}(\sigma_{2}+i\sigma_{1}))\\
S[\beta]=(-)^{n_{2}}\exp(\frac{\zeta}{2i}(\sigma_{2}+i\sigma_{1}))
       \end{array} \right\},  \label{eq:HOFS}
\end{equation}
where $n_{1}$ and $n_{2}$ take their values in $\{0,1\}$, and
$(\eta,\zeta)$ are homogeneous coordinates of ${\bf CP}^{1}
\approx S^{2}$.\footnote{The parametrization in ref.\cite{ezawa}
corresponds to the choice $n_{1}=n_{2}=0$ and $\eta=1$.}$^{,}$
\footnote{
One may claim that the following additional sectors exist:
\begin{eqnarray}
\check{\cal M}_{A1}&:&\quad S[\alpha]=\exp(\frac{\sigma_{1}}{2i}2\pi
),\quad S[\beta]=\exp\left(\frac{1}{2i}(A\sigma_{1}+B\sigma_{2})
\right)   \nonumber \\*
\check{\cal M}_{A2}&:&\quad\mbox{\it The same as }
\check{\cal M}_{A1}\mbox{ \it
with $\alpha$ and $\beta$ interchanged},
\end{eqnarray}
where $n\in{\bf Z}\setminus\{0\}$ and $B\in{\bf C}\setminus\{0\}$.
These sectors should probably be absorbed in $\check{\cal M}_{N}$
or in $\check{\cal M}_{F}^{n_{1}n_{2}}$ by taking an appropriate
gauge choice.}

To investigate the topology of $\check{\cal M}$, let us look for
flat SO(3,1) connections, i.e., $so(3,1)$ Lie algebra-valued
1-forms, which give a point on $\check{\cal M}$ as their holonomy.
By taking the equivalence classes under the
SL(2,{\bf C}) gauge transformations {\em which are homotopic
to the identity and thus are generated by six 1st class
constraints}, we find the representatives of each points on
$\check{\cal M}_{S}$
\begin{equation}
A=-\frac{\sigma_{1}}{2i}\{(u+i\alpha)dx+(v+i\beta)dy\}
\quad;
\label{eq:CSSS}
\end{equation}
and on $\check{\cal M}_{F}^{n_{1}n_{2}}$
\begin{equation}
A=-\frac{\sigma_{3}}{2i}2\pi(n_{1}dx+n_{2}dy)
-\frac{\sigma_{2}+i\sigma_{1}}{2i}e^{2\pi i(n_{1}x+n_{2}y)}
(\eta dx+\zeta dy),
\label{eq:CSFS}
\end{equation}
the range of the parameters are as before.

While we would like to extend $\check{\cal M}_{S}$ so that it
corresponds to the ADM phase space in a 1 to 1 fashion, it appears
that we cannot do so {\em at least we stick to the framework of CS
gauge theory}. It is because the gauge transformation
\begin{equation}
g=\exp\left(\frac{\sigma_{1}}{2i}(4\pi nx+4\pi my)\right)
\quad n,m\in {\bf Z}, \label{eq:LGT}
\end{equation}
is in fact homotopic to the identity in the space of SL(2,{\bf C})
gauge transformations and can be generated by six first class
constraints( at least in a certain limit). The parameters $u$ and
$v$ are defined modulo $4\pi$ and $\check{\cal M}_{S}$ has the
cotangent bundle structure ${\bf T}^{\ast}{\cal B}$ with its base
space ${\cal B}$ being an orbifold $T^{2}/{\bf Z}_{2}$.\footnote{
The base space and cotangent spaces are parametrized, respectively,
by $(u,v)$ and $(\alpha,\beta)$. ${\bf Z}_{2}$ is generated by
the inversion \cite{ezawa}.}
This $\check{\cal M}_{S}$ is precisely the phase space
of the SL(2,{\bf C}) CS gauge theory
which was given by Witten \cite{witten2}. However, the
spacetime having a point on $\check{\cal M}_{S}$ as their
holonomy are infinitely many. Let us consider the spacetimes whose
metric are
\begin{eqnarray}
ds^{2}&=&\frac{1}{\Lambda}[-dt^{2}+\cosh^{2}t\mbox{ }d\varphi^{2}
  +\sinh^{2}t\mbox{ }d\theta^{2}], \nonumber \\*
& &\mbox{ with }\left\{\begin{array}{l}
\varphi=(u+2\pi n)x+(v+2\pi m)y \quad(n,m\in{\bf Z})\\
\theta=\alpha x+\beta y  \quad, \end{array}\right.
\end{eqnarray}
where $(x,y)$ denotes a set of periodic coordinates with period 1.
All of these spacetimes have the same SO(3,1) holonomy
which is related with Eq.(\ref{eq:HOSS}).
This can be considered as a concrete realization of the
situation pointed out by Mess and Witten \cite{carlip},
that is, a holonomy map in the $\Lambda>0$ case
corresponds to an infinite, but discrete, set of non-diffeomorphic
spacetimes \cite{mess}\cite{witten2}. As Witten pointed out, to
specify a unique spacetime, we have to give additional \lq\lq
quantum numbers" $(n,m)$ which represent winding numbers around
nontrivial loops as well as a point on $\check{\cal M}_{S}$.

We are thus obliged to use a prescription by hand in order to
construct the phase space ${\cal M}$ of CSG which is related to
2+1 gravity more directly than $\check{\cal M}$.

First we construct the standard sector ${\cal M}_{S}$. We prepare
infinitely many copies of $\check{\cal M}_{S}$, each of which is
equipped with a set of \lq\lq quantum numbers" $(n_{1},n_{2})$.
We then cut the base space of each copies along three lines, each of
which links a conical singularity with one of the other three
conical singularities (Fig.1(a)(b)). By arranging the resultant
\lq\lq cotangent bundles" over triangles as is shown in Fig.1(c)
and gluing them together along the adjoining cuts, we obtain the
desired standard sector ${\cal M}_{S}$ (Fig.1(d)).
The base space of ${\cal M}_{S}$ is a cone
${\bf R}^{2}/{\bf Z}_{2}$ and is coordinatized by
\begin{equation}
\tilde{u}\equiv u+4\pi n_{1}\quad,\quad
\tilde{v}\equiv v+4\pi n_{2} \qquad
\biggl((\tilde{u},\tilde{v})\sim -(\tilde{u},\tilde{v})\biggr).
\end{equation}
We will henceforth call$(\tilde{u},\tilde{v})$ as $(u,v)$. We can
say that ${\cal M}_{S}$ is parametrized by connection
(\ref{eq:CSSS}) with $(u,v,\alpha,\beta)\in{\bf R}^{4}/{\bf Z}_{2}$.
As explained in the previous paper \cite{ezawa}, we can construct a
spacetime from a flat connection $A\in{\cal M}$ by taking an
appropriate (time-dependent) gauge transformation. In the CS
gauge theory, $u$($v$) and $u+4\pi$($v+4\pi$) are regarded to be
gauge equivalent. In general relativity, however,
differences of spacetimes themselves should be observable, so we
should be able to distinguish the $4\pi$ differences of $u$ and of
$v$. This justifies the use of ${\cal M}_{S}$ as a sector of the
physical phase space of CSG.

Next we consider the flat sector ${\cal M}_{F}$. Each
$\check{\cal M}_{F}^{n_{1}n_{2}}$ in the CS gauge theory
independently corresponds the space of special solutions in the ADM
formalism\cite{ezawa}. It turns out that the points on
$\check{\cal M}_{F}^{n_{1}n_{2}}$'s which are labeled by different
$(n_{1},n_{2})$ and which are parametrized by the same $(\eta,\zeta)
\in {\bf CP}^{1}$ give the same spacetime and the same SO(3,1)
holonomy.\footnote{These connections are indeed related
by a large local Lorentz transformation.}
Thus we cannot observe the difference in $(n_{1},n_{2})$
unless any fermionic observables exist.
Here we will take the viewpoint that we cannot distinguish different
$(n_{1},n_{2})$'s and we will regard all $\check{\cal M}_{F}
^{n_{1}n_{2}}$'s to be equivalent to $\check{\cal M}_{F}^{00}$,
which we will call ${\cal M}_{F}$.

The physical phase space ${\cal M}$ of CSG obtained by the above
prescriptions then the union of two
sectors ${\cal M}_{S}$ and ${\cal M}_{F}$ which have been given in
ref.\cite{ezawa}. The relation of ${\cal M}_{S}$ with the
ADM phase space and
the quantization of ${\cal M}_{S}$ can therefore be given in the
same form as the spacelike sector in the $\Lambda=0$ case
\cite{carli}\cite{ander}.\footnote{For reference, the phase space
$\check{\cal M}$ of the SO(3,1) Chern-Simons gauge theory is the
union of a cotangent bundle over an orbifold($\check{\cal M}_{S}$)
and four $S^{2}$'s ($\check{\cal M}_{F}^{n_{1}n_{2}}$).}

Next we look for new canonical coordinates which parametrize
${\cal M}_{S}$ and ${\cal M}_{F}$ in a unified form as in the
cases with $\Lambda\leq0$ \cite{louko}\cite{ezawa2}.
It is convenient to rewrite the parametrization (\ref{eq:CSSS})
(\ref{eq:CSFS}) as follows
\footnote{We have removed the origin of ${\cal M}_{S}$ which gives
a conical singularity (and a singular 1-dimensional universe).}
\begin{eqnarray}
{\cal M}_{S}\setminus\{(0,0)\}&:&
A=-\frac{\sigma_{1}}{2i}\sqrt{2\xi}(\cos\frac{\Theta}{2}e
^{\frac{i}{2}\phi}dx+\sin\frac{\Theta}{2}e^{-\frac{i}{2}\phi}dy)
\nonumber \\*
\quad with & & \xi\in{\bf C}\setminus\{0\},\quad \Theta\in[0,\pi],
\quad \phi\in[0,2\pi) ;\label{eq:CSSSU} \\
{\cal M}_{F} \quad\qquad &:&
A=-\frac{\sigma_{1}+i\sigma_{2}}{2i}(\cos\frac{\Theta}{2}e
^{\frac{i}{2}\phi}dx+\sin\frac{\Theta}{2}e^{-\frac{i}{2}\phi}dy)
\nonumber \\*
\quad with & &  \Theta\in[0,\pi],
\quad \phi\in[0,2\pi). \label{eq:CSFSU}
\end{eqnarray}
Then we introduce the following new parametrization:
\begin{eqnarray}
&& \qquad A^{new}=  \label{eq:CSPSU} \\
&&-\left[(\xi+\sqrt{\xi^{2}+1})^{1/2}\frac{\sigma_{1}}{2i}+
i(-\xi+\sqrt{\xi^{2}+1})^{1/2}\frac{\sigma_{2}}{2i}\right]
(\cos\frac{\Theta}{2}e
^{\frac{i}{2}\phi}dx+\sin\frac{\Theta}{2}e^{-\frac{i}{2}\phi}dy),
\nonumber
\end{eqnarray}
with $\xi\in{\bf C}$, $\Theta\in[0,\pi]$ and $\phi\in[0,2\pi)$.
This gives (\ref{eq:CSFSU}) for $\xi=0$. For $\xi\neq0$, we can
obtain (\ref{eq:CSSSU}) by performing a rigid gauge transformation
$g=\exp(\frac{\sigma_{3}}{2i}\psi)$ with $\psi\equiv\frac{1}{2i}
\ln\{(1+\sqrt{1+\xi^{2}})/\xi\}$. This \lq\lq unified phase space"
$$
{\cal M}^{\prime}\equiv{\cal M}_{S}\setminus\{(0,0)\}\oplus
{\cal M}_{F}
$$
is a Hausdorff manifold whose topology is ${\bf C}\times S^{2}$,
with $\xi$ and $(\Theta,\phi)$ parametrize ${\bf C}$ and $S^{2}$
respectively. This situation is different from those in
the cases with $\Lambda\leq0$, where the unified phase spaces are
non-Hausdorff\cite{louko}\cite{ezawa2}.\footnote{
If we unify $\check{\cal M}_{S}$ and $\check{\cal M}_{F}
^{n_{1}n_{2}}$, the resultant phase space may be non-Hausdorff.}

The symplectic structure of this unified phase space ${\cal M}
^{\prime}$ is given by
\begin{equation}
\omega=\frac{\sqrt{\Lambda}}{4}(dp_{\Theta}\wedge d\Theta +
dp_{\phi}\wedge d\phi), \label{eq:symp}
\end{equation}
where we have set $\xi\equiv p_{\phi}/\sin\Theta-ip_{\Theta}$.
This reproduces the symplectic structure of ${\cal M}_{S}$ and
explains that the symplectic structure of ${\cal M}_{F}$ vanishes.

To summarize, the reduced phase space $\check{\cal M}$ of the
SO(3,1) CS gauge theory is not suitable for describing 2+1 gravity
with positive cosmological constant. To describe the spacetime
smoothly, we have to construct the \lq\lq phase space ${\cal M}$ of
CSG" by considering the geometric feature of CSG and by making use
of the surgery which has been explained above. Such
prescription, however, cannot necessarily be natural. To find a more
natural procedure to construct ${\cal M}$, it will probably be
essential to consider the origin of the \lq\lq extra symmetry" Eq.
(\ref{eq:LGT}) which appears when the first order general relativity
passes to the SO(3,1) Chern-Simons gauge theory. If we can elucidate
this \lq\lq one to infinitely many correspondence" between SO(3,1)
CS gauge theoty and general relativity, the \lq\lq one to two
correspondence" between SO(2,2) CS theory and 2+1 gravity with
$\Lambda<0$ \cite{ezawa} will probably be made transparent since they
have the similar origin.

To complete the analysis of CSG (particularly on ${\bf R}\times
T^{2}$), some further issues remain
unresolved. We list a few of these problems:
i)  When we construct a spacetime from an SO(3,1) connection $A$
which is representative of a point on ${\cal M}$,
we have chosen a particular gauge. If we use another gauge, however,
we will probably obtain a different spacetime. The criterion for
the \lq\lq correct" choice of the gauge is therefore necessary; and
ii) in quantizing CSG on ${\bf R}\times T^{2}$, we have to take
into account the invariance under the modular transformations
\begin{equation}
S:(\alpha,\beta)\rightarrow(-\beta,\alpha),\quad
T:(\alpha,\beta)\rightarrow(\alpha+\beta,\beta),\label{eq:mod}
\end{equation} where $\alpha$ and $\beta$ are two generators of
$\pi_{1}(T^{2})$. There would be no problem when we quantize
${\cal M}_{S}$ alone since its description exactly coincides with
that of ${\cal M}_{s}$ in the $\Lambda=0$ case \cite{carli}
\cite{ander}. If we quantize the \lq\lq unified" phase space
${\cal M}^{\prime}$, however, to require the modular invariance
becomes a nontrivial task. Under (\ref{eq:mod}) the canonical
variables of ${\cal M}^{\prime}$ transform as follows:
\begin{eqnarray}
S &:&(\phi,\Theta,p_{\phi},p_{\Theta})\longrightarrow
(\pi-\phi,\pi-\Theta,-p_{\phi},-p_{\Theta})\quad,  \\
T &:&\left(\begin{array}{l}\phi \\ \Theta \\ p_{\phi} \\ p_{\Theta}
\end{array}\right)\longrightarrow
\left(\begin{array}{c}
-i\ln\left(\frac{\sin\frac{\Theta}{2}e^{i\phi}-\cos
\frac{\Theta}{2}}{\sqrt{1-\sin\Theta\cos\phi}}\right) \\
2\arctan\left(\frac{\sqrt{1-\sin\Theta\cos\phi}}{\cos
\frac{\Theta}{2}}\right) \\
\frac{\sin\frac{\Theta}{2}-\cos\frac{\Theta}{2}\cos\phi}{
\sin\frac{\Theta}{2}}p_{\phi}-2\sin\phi\cos^{2}\frac{\Theta}{2}
\mbox{ }p_{\Theta} \\
\frac{1+\cos^{2}\frac{\Theta}{2}-\sin\Theta\cos\phi}
{\sqrt{1-\sin\Theta\cos\phi}}\left\{\frac{\sin\phi}{2\sin
\frac{\Theta}{2}}p_{\phi}+(\sin\frac{\Theta}{2}-\cos\phi\cos
\frac{\Theta}{2})p_{\Theta}\right\}\end{array}\right).\nonumber
\end{eqnarray}
We can show by a straightforward calculation that the symplectic
structure (\ref{eq:symp}) is invariant under these transformations.
The quantum theory of ${\cal M}^{\prime}$ is therefore expected to
possess some symmetry under the modular group. To find the precise
modular covariance is left to the future investigation\cite{ezap}.

As for 3+1 gravity, it is possible that the holonomy
variables, which form a set of fundamental variables in Ashtekar's
formalism\cite{ashte}, do not give us sufficient information for
quantum gravity and that we have to add complementary quantum
numbers analogous to \lq\lq winding numbers" in the $SO(3,1)$ CSG.
It deserves further study whether this is indeed the case.

\vspace{.5in}

\noindent {\large Acknowledgments}

I would like to thank Prof. S. Carlip for useful advice and for his
informing me of some references on 2+1 gravity with positive
cosmological constant. I am also grateful to Prof. K. Kikkawa,
Prof. H. Itoyama and H. Kunitomo for helpful discussions and careful
readings of the manuscript.


\vspace{.5in}

{\LARGE Figure Captions}\\[.35in]
{\bf Fig.1 Surgery from $\check{\cal M}_{S}$ to ${\cal M}_{S}$.}\\
(a)The base space of $\check{\cal M}_{S}$. The four conical
singularities and three lines along which we cut are represented by
black marks and dotted lines, respectively.\\
(b) The result of cutting open.\\
(c) Arranging the infinitely many copies. Pay attention
the disposition of the quantum numbers $(n_{1},n_{2})$.\\
(d) The base space of ${\cal M}_{S}$, which is a cone with
deficit angle $\pi$.

\end{document}